\documentclass[12pt]{article}

\newcommand{\beqn}{\begin{eqnarray}}
\newcommand{\eeqn}{\end{eqnarray}}
\newcommand{\Z}{{Z \!\!\! Z}}
\newcommand{\dd}{{\mbox d}}
\newcommand{\cL}{{\cal L}}
\newcommand{\eq}[1]{(\ref{#1})}

\newcommand{\itep}
{~\vspace{-1.5cm}
\begin{flushright}
{\large ITEP-TH-28/99}
\end{flushright}
\vspace{1.0cm}}

\begin{document}
\baselineskip=14pt
\begin{center}

\itep

{\large\bf Classical String Solutions}
\vskip 0.2cm
{\large \bf in Effective Infrared Theory of
$\mathbf{SU(3)}$ Gluodynamics}

\vskip 1.0cm
{\bf M.N.~Chernodub}
\vskip 4mm
{\it Institute of Theoretical and Experimental Physics},\\
{\it B.Cheremushkinskaya 25, Moscow, 117259, Russia}

\end{center}

\begin{abstract} We investigate string solutions to the classical
equations of motion ("classical QCD strings") for a dual Ginzburg-Landau
model corresponding to $SU(3)$ gluodynamics in an abelian projection. For
a certain relation between couplings of the model the string solutions are
defined by first order differential equations. These solutions are related
to vortex configurations of the Abelian Higgs model in the Bogomol'ny
limit. An analytic expression for the string tension is derived and the
string--string interactions are discussed. Our results imply that
the vacuum of $SU(3)$ gluodynamics is near a border between 
type-I and type-II dual superconductivity. 
\end{abstract}

\vskip 0.3cm
{\bf 1.}
Last years the problem of color confinement in $SU(N)$ Yang--Mills
theories has been intensively studied in the framework of the abelian
projection approach~\cite{tHo81}. This approach is based on a partial
gauge fixing which reduces the gauge symmetry from non--abelian gauge
group to its (maximal) abelian subgroup. The diagonal elements of the
gluon field transform under residual gauge transformations as abelian
gauge fields, while the off-diagonal elements transform as abelian
matter vector fields.  Due to compactness of the Yang--Mills gauge
group the residual abelian subgroup is also compact. This leads to
appearance of abelian monopoles in the abelian gauge. According to
the dual superconductor scenario~\cite{tHoMa76} the color confinement
may be explained at the classical level: if monopoles are
condensed then a string forms between color charges. This string is
an analog of the Abrikosov string~\cite{Abr} in a superconductor and
the abelian monopoles are playing the role of the Cooper pairs.

The dual superconductor mechanism has been confirmed by numerous lattice
simulations of $SU(2)$ Yang--Mills theory~\cite{RecentReviews}. Moreover,
the vacuum of $SU(2)$ Yang--Mills theory in the abelian projection was
shown~\cite{TypeI-II} to be close to the border between type--I and
type--II dual superconductors (masses of the monopole and the dual gauge
boson are approximately the same). There are also strong numerical
indications that the vacuum of $SU(3)$ gluodynamics exhibits dual
superconductor properties~\cite{SU3-DualSuperconductor}. Analytical
investigations~\cite{BaBaZa85} of the string configurations in a
non--abelian dual superconductor model of $SU(3)$ gluodynamics shows that
the dual vacuum is close to the border between type-I and type-II
superconductivity. In Ref.~\cite{Baetal98} similar conclusion has been
derived where the $SU(3)$ confining string was related 
to the Abrikosov--Nielsen--Olesen classical string solution~\cite{Abr,NiOl}
in the $U(1)$ Higgs model.

In this paper we investigate an effective $abelian$ model of $SU(3)$ 
gluodynamics suggested in Ref.~\cite{MaSu81}. This model is based on dual
superconductivity conjecture which is supposed to describe infrared
properties of the $SU(3)$ gluodynamics vacuum. An essential difference
between $SU(2)$ and $SU(3)$ Yang--Mills theories in an abelian projection
is the presence of two independent string configurations in the later. The
existence of two string types guarantees the colour neutrality of the
quark bound states~\cite{KoCh98}. A possible string representation of
this model in a regime of extreme type-II superconductivity (the monopole
mass is infinite) has been discussed in Refs.~\cite{KoCh98,An98}.
The classical string configurations in the static baryon have been 
studied numerically in Refs.~\cite{Baryon,MaMaSu90}.
The results of Ref.~\cite{MaMaSu90} 
suggest that the vacuum of $SU(3)$ gluodynamics lies 
near a border between type--I and type--2 superconductivity. 
Below we study analytically the confining string regarding 
the $SU(3)$ gluodynamics as a dual superconductor 
with finite masses of the monopole and dual gauge
boson fields. A special attention is payed to the border between 
different types of superconductivity.

\vskip 0.3cm
{\bf 2.}
The Lagrangian of ${[U(1)]}^2$ Higgs model corresponding to $SU(3)$
gluodynamics is\footnote{In this paper the summation over
the Latin indices $i$ and $j$ is taken only if explicitly 
indicated.}~\cite{MaSu81}:
\beqn
\cL = - \frac{1}{4} F^a_{\mu\nu} F^{a,\mu\nu} +
\sum\limits^3_{i=1} \Bigl[\frac{1}{2}
{\bigl| D^{(i)}_\mu \chi_i\bigr|}^2 - \lambda
{\Bigl({|\chi_i|}^2 - \eta^2_i \Bigr)}^2\Bigr]\,,
\label{Lagrangian}
\eeqn
where $F^a_{\mu\nu} = \partial_\mu B^a_\nu - \partial_\nu B^a_\mu$ is
the field strength for the gauge fields $B^a_\mu$, $a=3,8$,
$D^{(i)}_\mu = \partial_\mu + i g \varepsilon^a_i B^a_\mu$ is the
covariant derivative acting on the Higgs fields $\chi_i$, $i=1,2,3$.
The $\epsilon$'s are the root vectors of the group $SU(3)$:  $\vec
\epsilon_1=(1, 0)\,,\vec \epsilon_2=(-1 \slash 2, -{\sqrt{3} \slash
2})\,,\vec \epsilon_3=(-{1 \slash 2}, {\sqrt{3} \slash 2})$.

The gauge fields $B^{3,8}_\mu$ are dual to the diagonal components
$a=3,8$ of the gluon field $A^a_\mu$. The Higgs fields $\chi_i$ are
associated with the monopole degrees of freedom which appear due to
compactness of the residual abelian gauge group, ${[U(1)]}^2$, in an
abelian projection $SU(3) \to {[U(1)]}^2$. Lagrangian \eq{Lagrangian}
respects the dual $[U(1)]^2$ gauge invariance:  $B^a_\mu \to B^a_\mu
+ \partial_\mu \alpha^a$, $\theta_i \to \theta_i + g (\varepsilon^3_i
\alpha^3 + \varepsilon^8_i \alpha^8) \,\, \mbox{mod} \, 2 \pi$,
$a=3,8$, $i =1,2,3$, where $\alpha^3$ and $\alpha^8$ are the
parameters of the gauge transformation. The phases of the Higgs fields 
satisfy the following relation:
\beqn
\sum\limits^3_{i=1} \arg \chi_i =0\,.
\label{phases}
\eeqn

The model admits vortex--like solutions to the classical equations of
motion similar to the Abrikosov vortex configurations in the
Ginzburg--Landau model of superconductivity. The vortices carry
quantized fluxes of electric fields since the matter fields $\chi_i$
correspond to the magnetically charged particles. We choose the
following anzatz for the field configuration of the static straight
vortex parallel to the $z$--axis:
\beqn
\begin{array}{lcll}
\chi_i & = & \eta_i \, f_i (\rho) e^{- i n_i \varphi} \,,
& \quad n_i \in \Z\,,\quad i = 1,2,3\,,\\
\vec{B}^a & = & - \frac{1}{g \rho} v^a(\rho)
\cdot \vec{e}_\varphi\,, & \quad a=3,8\,.\\
\end{array}
\label{anzatz}
\eeqn
where we have used the standard notations for the cylindrical
coordinates $(\varphi,\rho,z,t)$ in four dimensional space. Note that
the vortex configuration does not depend on $z$-- and
$t$--coordinates.

The finite energy configuration must satisfy the condition
$D^{(i)}_\mu \chi_i = 0$ at spatial infinity. This condition
implies the quantization of the total flux $\Phi^a = \int \! \dd x \dd y
\, H^a$ of the vortex magnetic field $H^a = F^a_{12}$:
\beqn
\Phi_i \equiv \sum\limits_{a=3,8} \varepsilon^a_i \, \Phi^a = \frac{2
\pi n_i}{g}\quad\mbox{or}\quad \Phi^a = \frac{4 \pi}{3 g}
\sum^3_{i=1} \varepsilon^a_i n_i\,,
\label{quant}
\eeqn
where we have used the relation $\sum^3_{i=1} \epsilon^a_i
\epsilon^b_i = 3\delta^{ab} \slash 2$. In particular this condition
implies that the strings must carry both components, $\Phi^3$ and
$\Phi^8$, of the fluxes.

According to eq.\eq{phases} the integer winding numbers $n_i$ satisfy
the following relation:
\beqn
\sum\limits^3_{i=1} n_i = 0\,.
\label{n}
\eeqn

\vskip 0.3cm
{\bf 3.} To analyse the vortex solutions we begin with the Bogomol'ny
method~\cite{Bo76}. Consider the energy density per unit length
of the string (string tension):
\beqn
\sigma = \int \! \dd^2 x \, \Bigl[ \frac{1}{2} {(H^a)}^2 +
\sum\limits^3_{i=1} \Bigl(\frac{1}{2} \sum\limits_{\alpha=1,2}
{\bigl|D^{(i)}_\alpha \chi_i \bigr|}^2
+ \lambda {\Bigl( {|\chi_i|}^2 - \eta^2_i\Bigr)}^2\Bigr)\Bigr]\,.
\label{star}
\eeqn
The first term in this equation can be rewritten as follows:
$\sum^2_{a=1} {(H^a)}^2 \slash 2 = \sum^3_{i=1} {H_i}^2 \slash 3$,
where $H_i = \sum^2_{a=1} \epsilon^a_i H^a$.
The second term in eq.\eq{star} can be represented as follows:
\beqn
\sum\limits_{\alpha=1,2} {\bigl|D^{(i)}_\alpha \chi_i\bigr|}^2 =
 {\Bigl| \Bigl(D^{(i)}_1 \pm i D^{(i)}_2 \Bigr)\chi_i\Bigr|}^2
\mp 2 \varepsilon_{\alpha\beta} \partial_\alpha J^{(i)}_\beta \pm
i \chi^+_i \Bigl[D^{(i)}_1,D^{(i)}_2 \Bigr] \chi_i\,,
\label{twostars}
\eeqn
where the current
$J^{(i)}_\alpha = i \chi^+_i D^{(i)}_\alpha \chi_i$
vanishes faster than $\rho^{-1}$ as $\rho \to \infty$ for the
configuration with a finite energy. Therefore the second
term in the {\it r.h.s.} of eq.\eq{twostars} gives zero
contribution to the string tension \eq{star}. The last term in
eq.\eq{twostars} can be simplified: $i \chi^+_i
[D^{(i)}_1,D^{(i)}_2 ]  \chi_i= - g H_i \, {|\chi_i|}^2$.
Thus the string tension \eq{star} can be expressed in the form:
\beqn
\sigma & = & \int \dd^2 x \, \sum\limits^3_{i=1} \Bigl[
\frac{1}{2} {\Bigl|\Bigr(D^{(i)}_1 \pm i D^{(i)}_2 \Bigr)
\chi_i\Bigr|}^2 + \frac{1}{3} H^2_i \mp \frac{g}{2}
H_i \, {|\chi_i|}^2 +
\lambda {\Bigl( {|\chi_i|}^2 - \eta^2_i \Bigr)}^2\Bigr]
\nonumber\\
& = & \int \dd^2 x \, \sum\limits^3_{i=1} \Bigl[\frac{1}{2}
{\Bigl|\Bigr(D^{(i)}_1 \pm i D^{(i)}_2 \Bigr)
\chi_i\Bigr|}^2 +  \frac{1}{3} {\Bigl(H_i \mp \frac{3 g}{4}
({|\chi_i|}^2 - \eta^2_i)\Bigl)}^2
\label{act}\\
&  & \pm \frac{g}{2} \eta^2_i \Phi_i
+ \Bigl(\lambda - \frac{3 g^2}{16}\Bigr)\,
{\Bigl( {|\chi_i|}^2 - \eta^2_i \Bigr)}^2\Bigr]\,,
\nonumber
\eeqn
where the magnetic flux $\Phi_i$ is quantized according to
eq.\eq{quant}. The sign in front of the flux term is chosen so that
the contribution of this term is positive.

The analogue of the Bogomol'ny limit~\cite{Bo76} is defined
by the following condition:
\beqn
\lambda = 3 g^2 \slash 16\,,
\label{BL}
\eeqn
which guarantees that the potential term in eq.\eq{act} vanishes.
Contrary to the case of Abelian Higgs model this condition does not
imply an equivalence of the vector and scalar boson masses. The
equivalence holds only in the physically relevant case when
the vacuum expectation values of the Higgs monopole fields $\chi_i$ 
are degenerate, $\eta_i \equiv \eta_j$, $i,j=1,2,3$ (the color 
symmetry is unbroken).

In this case the vortex configuration is defined by the first order
field equations:
\beqn
\Bigl(D^{(i)}_1 \pm i D^{(i)}_2 \Bigr)  \chi_i = 0\,; \quad
H_i \mp \frac{3 g}{4} \Bigl({|\chi_i|}^2 - \eta^2_i
\Bigr) = 0\,;\quad i =1,2,3\,,
\label{EOM}
\eeqn
or, for anzatz \eq{anzatz},
\beqn
f_i'(\rho) \pm (v_i(\rho) - n_i) f_i(\rho) \rho^{-1} = 0\,,\quad
\pm v_i'(\rho) + g^2 \eta^2_i (f^2_i (\rho) - 1) \rho = 0\,,
\label{anzeqs}
\eeqn
where $v_i = \sum_{a = 3,8} \varepsilon^a_i v^a$ and the prime
denotes the differentiation with respect to $\rho$. The system of
equations of motion \eq{anzeqs} is over-defined and therefore a
classical vortex solution to these equations does not exist in
general case. However we will show below that for a physical case
the classical solution can be found.

According to eqs.(\ref{quant},\ref{act}) the solutions to
the equations of motion \eq{EOM} or \eq{anzeqs}
saturate an analog of the Bogomol'ny bound for the string tension:
\beqn
\sigma = \pi \sum\limits^3_{i=1} |n_i| \eta^2_i\,.
\label{energy}
\eeqn
Thus parallel strings which carry the same flux do not interact with
each other since energy of $N$ strings with the same flux $\Phi^a$ is
the same as the energy of the single string with the flux $N \Phi^a$.
This is a general property of classical solutions in the Bogomol'ny
limit. The vortices start to interact when relation \eq{BL} is not
satisfied. A similar situation happens on a border between type-I and 
type-II superconductors.

\vskip 0.3cm
{\bf 4.} The
$SU(3)$ string in an abelian projection may be considered as a
composite~\cite{KoCh98,An98} of the three {\it elementary} strings
with the winding numbers $n_i$ of the Higgs fields $\chi_i$ subjected
to relation~\eq{n}. The phase of the Higgs field must be singular at
the center of the elementary string with non-zero winding number
$n_i$. Since all the Higgs field phases are subjected to the
condition \eq{phases} the center of the elementary strings 
must coincide. The tension of the $SU(3)$ string
depends on the fluxes $\Phi^a$ carried by the string and the fluxes
are being related to the winding numbers $n_i$ in accordance with the
quantization condition \eq{quant}.

Consider the classical field configuration for the abelian counterpart
of the $SU(3)$ string in the Bogomol'ny limit. The lowest energy
string configuration in the neutral quark--anti-quark system has the
winding numbers $1$, $-1$ and $0$ (there are three
string configurations corresponding to different colors of the
quarks, or, equivalently, to different permutations of the $n$'s).
Note that the same winding numbers are carried by segments of the $SU(3)$ 
classical string configuration in the baryon~\cite{KoCh98}.

For the sake of
generality we consider below the classical string solution with the
winding numbers $n$, $-n$ and $0$.  The solution to the equations of
motion \eq{anzeqs} is:
\beqn
f_i(\rho) = f^{\mathrm{ANO}}_{|n_i|} (\rho,
g \eta_i) \,,\quad
v^a = \frac{2}{3} \sum\limits^3_{i=1} {\mathrm{sign}}(n_i)\,
\varepsilon^a_i \, v^{\mathrm{ANO}}_{|n_i|} (\rho, g \eta_i)\,,
\label{SB}
\eeqn
where $f^{\mathrm{ANO}}_n(\rho,m)$ and $v^{\mathrm{ANO}}_n(\rho,m)$
are the characteristic functions of the Abrikosov--Nielsen--Olesen
(ANO) vortex solution~\cite{Abr,NiOl} in the Bogomol'ny limit of the
Abelian Higgs model (AHM). These functions are defined by the
first order Bogomol'ny equations in the AHM~\cite{VeSc76}:
\beqn
f^{\mathrm{ANO}'}_n \pm (v^{\mathrm{ANO}}_n - n)
f^{\mathrm{ANO}}_n \rho^{-1} = 0\,,\quad
\pm v^{\mathrm{ANO}'}_n(\rho) + m^2
\Bigl({(f^{\mathrm{ANO}}_n)}^2 - 1\Bigr) \rho = 0\,,\quad n>0 \,.
\eeqn
The functions $f^{\mathrm{ANO}}_n$ and $v^{\mathrm{ANO}}_n$ have been
determined numerically in Ref.~\cite{VeSc76}. Note that
$f^{\mathrm{ANO}}_0=1$ and $v^{\mathrm{ANO}}_0=0$.

In the degenerate case $\eta_i = \eta$, $i=1,2,3$, the characteristic
functions $v^a$ for the string with the winding numbers $n$, $-n$ and
$0$ have a simple form:
\beqn
v^a (\rho) = \frac{\Phi^a}{2 \pi n} \,
v^{\mathrm{ANO}}_{n} (\rho, g \eta)\,,\quad n>0\,,
\eeqn
where the fluxes $\Phi^a$ are given by eq.\eq{quant} and 
according to eq.\eq{energy} the string tension is 
\beqn
\sigma_n = 2 \pi n \eta^2\,.
\label{sigma}
\eeqn

The solutions for a general case ($n_i \neq 0$, $i=1,2,3$) are unlikely to
exist due the fact that system \eq{anzeqs} is over-defined. Indeed
according to the definition of the vectors $v_i$ we need the
fulfillment of the following condition: $\sum^3_{i=1} n_i \,
v^{\mathrm{ANO}}_{|n_i|} (\rho, g \eta_i) = 0$ with given vorticities
$n_i$ and {\it integer} parameters $\eta_i$. Since the functions
$v^{\mathrm{ANO}}_n$ are solutions of the nonlinear equations the
linear dependence of these functions would be unnatural. Thus the
Bogomol'ny bound \eq{energy} for the string tension is likely to be
reached only for the strings composed of two elementary abelian strings 
(the third vorticity component must be zero).

\vskip 0.3cm
{\bf 5.} The interaction between vortices is defined by their Higgs
and gauge boson field profiles. The Higgs (gauge) boson mediated
interaction between two parallel identical vortices is attractive
(repulsive).  In the Bogomol'ny limit the Higgs and gauge boson
fields acts in the same range since at large distances the
profile functions $v^{\mathrm{ANO}}(\rho,m_1)$ and
$f^{\mathrm{ANO}}(\rho,m_2)$ behave like $e^{- m_1 \rho}$ and 
$e^{- m_2 \rho}$, respectively. According to eq.\eq{SB} the mass
$m_1$ equals to the mass $m_2$ in the  Bogomol'ny limit.
Thus the interactions compensate each
other and the net force between vortices is zero\footnote{Note that 
the ${[U(1)]}^2$ vortices are not interacting for all vortex--vortex 
separations contrary to, {\it e.g.}, $\Z_3$ vortices (these are 
also supposed~\cite{Va96} to be relevant to confinement in QCD.). 
The potential between $\Z_3$ vortices may be attractive at short 
separations and repulsive at large distances~\cite{HeVa98}.}.

When $\lambda < 3 g^2 \slash 16$ the vortices are
attracting since the attraction due to exchange of the Higgs boson
prevails over the gauge boson attraction. In this region of the
parameter space the strings with identical fluxes tend to join. For
$\lambda > 3 g^2 \slash 16$ the strings repel each other and therefore the
strings with multiple vorticities are unstable: they tend to decay
on strings with smallest vorticities.

Using similar arguments one can show that the strings with different
fluxes (winding numbers) are always attractive in the type-I region. 
In the type-II region the gauge--mediated force between strings $A$ and
$B$ is  proportional to $\sum_{a=1,2} \Phi^a_A \Phi^a_B \propto 
\sum^3_{i=1} n^A_i n^B_i$. The interaction between the stings is
attractive if this number is negative and repulsive otherwise.

In an ordinary type--I (type--II) superconductor the Abrikosov
vortices attract (repel) each other and the Bogomol'ny limit is a
border line between the two types. By analogy, relation \eq{BL}
defines the border line between different types of the dual
superconductivities.

The closeness of the $SU(3)$ gluodynamics vacuum to the border between
type-I and type-II may be checked with the help of eq.\eq{sigma}
which relates vacuum expectation values of the monopole fields $\eta$ 
and the string tension {\it at the border}. Using the
estimate~\cite{MaSu81,Baryon,MaMaSu90,Hideo} for the vacuum expectation
value of the Higgs fields,
$\eta \approx 175$~MeV, we get\footnote{Note that the notations of 
Refs.~\cite{MaSu81,Baryon,MaMaSu90,Hideo} differ from our
notations.} the value of the tension of the string with the lowest flux,
$\sigma \equiv \sigma_1 \approx (440\,\,\mbox{MeV})^2$ which is quite
close to a phenomenological value. Thus the gluodynamics vacuum is likely 
to be near the border between type-I and type-II dual superconductivity.

\section*{Conclusions} The effective model of $SU(3)$ gluodynamics in
an abelian projection possesses classical vortex solutions ("QCD
strings"). For a certain relation between couplings of this model ($16
\lambda = 3 g^2$) the vortex solutions are related to the solutions of the
Abelian Higgs model in the Bogomol'ny limit. This relation also defines the
border between type-I and type-II dual superconductivity. The string
tension in the Bogomol'ny limit is quantized according to eq.\eq{energy}.
Using this formula and the phenomenological value of the vacuum
expectation value of the monopole field we show that the 
vacuum of the $SU(3)$ gluodynamics lies near the border between type--I 
and type--II dual superconductivity.

\section*{Acknowledgments}
The author is grateful to M.I.~Polikarpov, T.~Suzuki and T.~Vachaspati
for useful discussions.
The author feel much obliged for the kind hospitality extended to him
at the Theoretical Physics Institute at the University of Minnesota
where this work has been started.  This work was supported by the
grants INTAS-96-370, RFBR-96-15-96740, RFBR-99-01-01230 and INTAS Grant
96-0457 within the research program of the International Center for
Fundamental Physics in Moscow.



\end{document}